\documentclass{BCTagents}
\usepackage{times}
\usepackage{graphicx}
\usepackage{latexsym}
 \newcommand{\q}[1]{``#1''}

\begin{document}

\title{Modelling Behaviour Change using Cognitive Agent Simulations}

\author{Catriona M. Kennedy \institute{School of Computer Science, University of Birmingham, UK,
email: catm.kennedy@gmail.com} }

\maketitle
\bibliographystyle{BCTagents}

\begin{abstract}
In health psychology, Behaviour Change Theories (BCTs) play an important role in modelling human goal achievement in adverse environments. Some of these theories use concepts that are also used in computational modelling of cognition and affect in AI. Examples include dual-process architecture and models of motivation.  It is therefore important to ask whether some BCTs can be computationally implemented as cognitive agents in a way that builds on existing AI research in cognitive architecture.

This paper presents work-in-progress research to apply selected behaviour change theories to simulated agents, so that an agent is acting according to the theory while attempting to complete a task in a challenging scenario. Two behaviour change theories are selected as examples (CEOS and PRIME). The research is focusing on complex agent architectures required for self-determined goal achievement in adverse circumstances where the action is difficult to maintain (e.g. healthy eating at office parties). 

Such simulations are useful because they can provide new insights into human behaviour change and improve conceptual precision. In addition, they can act as a rapid-prototyping environment for technology development. High-level descriptive simulations also provide an opportunity for transparency and participatory design, which is important for user ownership of the behaviour change process.
\end{abstract}

\section{INTRODUCTION}
Behaviour change theories (BCTs) play an important role in the modelling of human goal achievement in adverse circumstances. Such theories can also inform the design of technology to support behaviour change. Examples include context-sensitive messages and self-monitoring (e.g. for physical activity). In parallel with these developments, cognitive and affective models in AI (e.g. \cite{Sloman05:arch}, \cite{hudlicka07:reasons}) have been used to simulate motivation and emotion. Some behaviour change theories use concepts that are similar to those used in AI models. These include the idea of a \q{dual-process} architecture where there is a contrast between a fast, reactive layer and a slower, more deliberative layer involved in explicit reasoning. 

This paper presents work-in-progress research to apply selected behaviour change theories to simulated cognitive agents, so that an agent is acting according to the theory. In other words, the agent attempts to achieve a goal in accordance with the behaviour change theory in a simulated scenario in which difficulties arise. Such simulations can help to improve understanding of human motivation and provide a rapid prototyping environment for technology design. In particular, a simulated autonomous agent may act as a blueprint for an assistant agent in the real world, which makes recommendations instead of executing decisions itself. 

The type of behaviour change under consideration has the following properties: (a) it is self-determined; (b) it is difficult to achieve and maintain in circumstances where opposing motives are activated (e.g. stop smoking) (c) it is not limited to health, but involves any situation where actions should be consistent with values (e.g. resisting sales pressure). \q{Self-determined} means that a person has a goal that they have spent time considering, and they have \q{ownership} of this goal. For example, such goals may arise from their principles and values (e.g. support the environment) or from the desire to improve their health (e.g. be a non-smoker). In contrast, non-self-determined goals and actions are externally specified or suggested. They may be reward-based (extrinsically motivated) or they may arise rapidly in response to a perceived need (e.g. get a product before it sells out). The concept of self-determination is elaborated in Self-determination Theory \cite{Ryan02:SDT}. 

Since the behaviour change is self-determined, it is important that users have ownership and control over the design of assistance technology. In addition to transparent design, the assistant agent should be able to explain the reasons for its messages, decision recommendations or any other actions that it takes. For this purpose, we are focusing on symbolic reasoning and knowledge-based approaches to AI. At a later stage of development, symbolic methods may be combined with other paradigms if appropriate (e.g neural networks). 

The remainder of this paper is structured as follows: Section \ref{related} outlines related work. Section \ref{BCTs} summarises two representative behaviour change theories (CEOS and PRIME). Section \ref{cog} outlines a particular approach to cognitive agents which we are using as a foundation. Section \ref{shared} identifies connections between these different models. Sections \ref{sim} and \ref{prot} present our current work in combining BCT concepts with cognitive agents. Finally, section \ref{challenges} discusses research challenges and future work.

\section{RELATED WORK}\label{related}

Our work is related to symbolic cognitive architectures which model human cognition and affect. These include MAMID \cite{hudlicka07:reasons}, H-CogAff \cite{Sloman05:arch} and EMA \cite{Marsella07:EMA}. MAMID  models the effects of emotion on decision-making and has been used to model affective disorders \cite{hudlicka17:healthcare}.  H-CogAff provides a generic model for human cognition and affect, emphasising the interaction between fast reactive processes and slower, deliberative processes. EMA \cite{Marsella07:EMA} is a computational model of emotion based on appraisal theory (as presented in e.g. \cite{Scherer99:appraisal}) and includes coping mechanisms for managing negative emotions, a capability that is particularly relevant for behaviour change. The \q{Emotion Machine} architecture of Minsky \cite{minsky07:book} is a general distributed framework, involving \q{critics}, which provide different methods for evaluating a situation and \q{selectors}, which are strategies for determining action.  Models of cognitive control are also important for self-determined goals. For example, the ARCADIA project \cite{bello17:agency} introduces different levels of cognitive control, where the highest level is necessary for recognising and keeping {\it commitments}, which also requires the control of attention.

Serious games \cite{Schuller13:pervasive},\cite{Fleming17:serious} are a potential format for behaviour change assistance and many of them address the same problems that agent-based simulations have to solve (e.g. realistic scenarios and visualisation). Furthermore, serious games can include agents with simulated emotions and coping strategies (e.g. \cite{Broekens16:appraisal}, \cite{Popescu14:GAM}).

The Domino model \cite{Fox13:canonical} is a cognitive agent model for decision-making, which is intended to be applied to decision support, with an emphasis on knowledge-based approaches. This is relevant to our work because we are aiming for an assistant agent that may take the form of a decision support system (although not limited to this format). Domino agents can be autonomous or they can be assistant agents which can guide a user through recommended actions and plans, where some of the recommendations depend on dynamic events. Other approaches to knowledge-based decision support are surveyed in \cite{Liu14:KBS}.

\section{SELECTED BEHAVIOUR CHANGE THEORIES}\label{BCTs}

We are selecting BCTs that address the following topics:
\begin{itemize}
\item 
Behaviour change that is difficult to achieve or maintain due to circumstances that activate opposing motives or emotions.
\item
Intrinsic motivation - self-determined goals.
\item
Cognitively rich models - emphasising reasoning and planning combined with impulsive actions.
\end{itemize}
We are currently looking at PRIME theory \cite{West07:PRIME} and CEOS \cite{Borland17:CEOS}. Other relevant BCTs include self-determination theory \cite{Ryan02:SDT}, reflective impulsive model (RIM) \cite{Strack04:RIM}, implementation intentions \cite{Gollwitzer99:imp}, control theory \cite{Carver82:control} and self-regulation theory \cite{Bandura91:SRT}. A comprehensive review is given in \cite{Michie14:ABC}.

\subsection{PRIME theory of motivation}
PRIME \cite{West07:PRIME} stands for \q{Plans, Responses, Impulses, Motives, Evaluations}, each of which is defined below:
\begin{itemize}
    \item Plans - intentions: prepare actions in advance
    \item Evaluations - beliefs about what is good or bad
    \item Motives - wants or needs
    \item Impulses - readiness or tendency for action
    \item Responses - action
\end{itemize}
Plans and evaluations are at the highest level and they are usually considered in advance. These in turn generate motives (e.g. wanting to stop smoking). Motives generate readiness for action (impulses), which in turn generate action. \\ \\
\noindent
PRIME has three principles:
\begin{enumerate}
\item 
Wants and needs at each moment drive behaviour.
\item 
Beliefs/intentions about good or bad must produce sufficiently strong needs at the moment of action.
\item
A sense of identity is potentially a strong source of wants/needs that can override drives that are activated in a particular moment.
\end{enumerate} 
In the early stages, we are focusing on (1) and (2), although a sense of identity is an important consideration in the longer term. The generation of sufficiently strong needs at the moment of action is an important principle that will be discussed in detail later.

\subsection{CEOS theory}
CEOS \cite{Borland17:CEOS} stands for \q{Context, Executive and Operational Systems}. The main principle is that human behaviour is directed and implemented like an organisation, which involves two levels:
\begin{itemize}
\item 
Operational System (OS): generates all behaviour and responds to needs as they occur.
\item
Executive System (ES): directs the OS, and is responsible for goal-setting and planning.
\end{itemize}
The OS responds only to actual events as they are experienced, while the ES can imagine non-actual states and negations (e.g. hypothetical events in the future). ES plans cannot be realised without sufficient activation of the OS.

\section{COGNITIVE AGENTS}\label{cog}

As a generic example of cognitive agent architectures, we have selected a simplified version of H-CogAff \cite{Sloman05:arch}, which means \q{Human Cognition and Affect}. This architecture is made up of three layers: reactive, deliberative and metacognitive. The reactive layer responds quickly to actual events. The deliberative layer is slower and has a sequential nature, in the sense that a \q{current process} exists (the current focus of attention). An important capability of the deliberative layer is that it can generate non-actual, \q{what-if} events (hypothetical reasoning). Metacognition is a reflective layer which can monitor and control the deliberative and reactive layers. Deliberation and metacognition working together can be defined as \q{executive control}. 

\begin{figure}
\includegraphics[width=8cm]{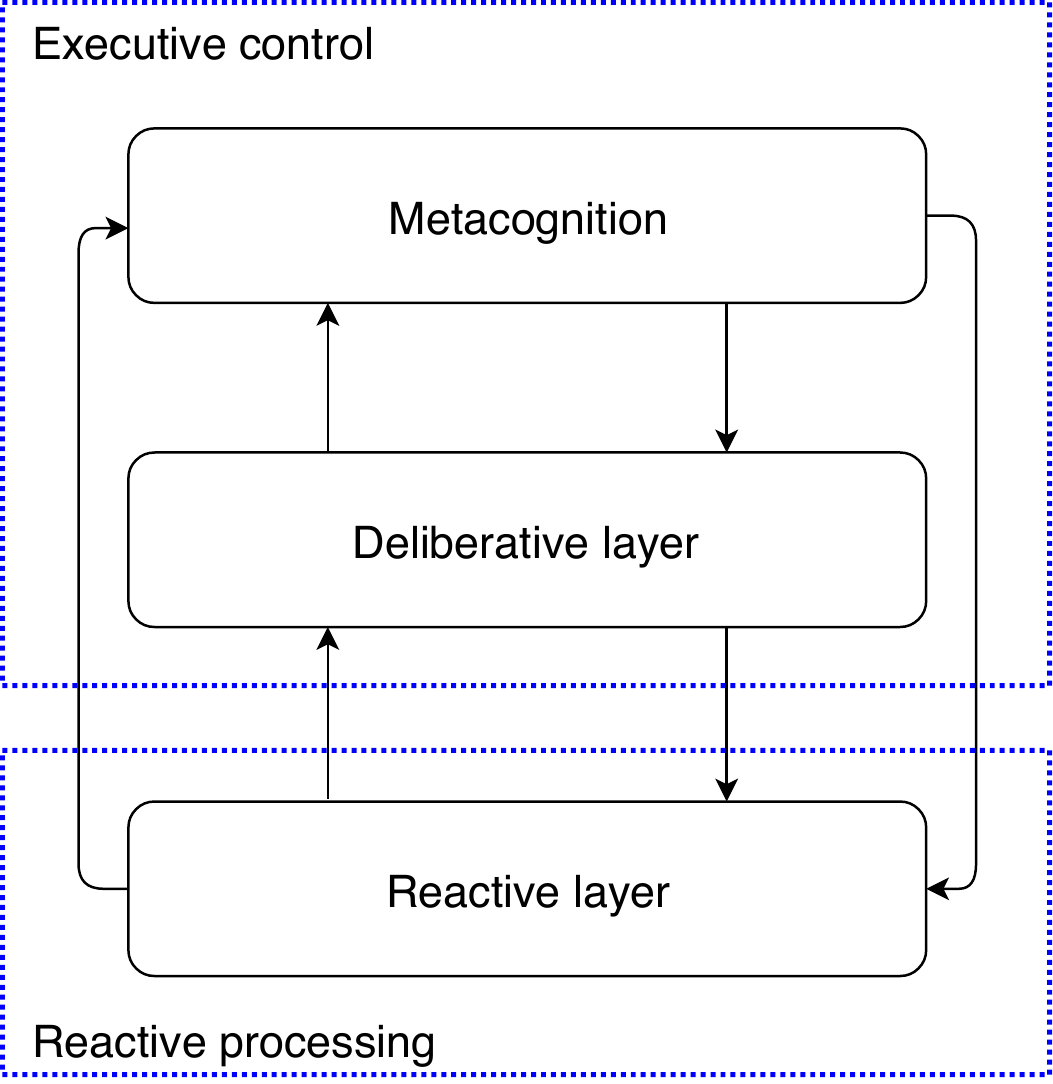}
\caption{A simplified version of H-CogAff} \label{fig1}
\end{figure}

The simplified H-CogAff architecture is shown schematically in Figure \ref{fig1}. It does not show the complexities of the original H-CogAff. For example, the details of multi-level perception and action are omitted, as well as the agent environment.
H-CogAff is a representative example of this kind of multi-layered architecture, and not the only \q{right} solution. There are others with similar concepts (e.g. MAMID \cite{hudlicka07:reasons}). 

\subsection{Metacognition}

Metacognition is often called \q{thinking about thinking} \cite{cox11:meta-intro}. It is made up of two processes: monitoring and control. Monitoring evaluates the performance of the deliberative and reactive layers; control makes necessary adjustments to the processing in these layers. For example, metacognition may detect a gap in the agent's knowledge due to an observed outcome being different from the expected one. In this case, it may initiate a plan to learn the required knowledge (e.g. \cite{cox99:multistrategy}). 

\subsection{Affective states and emotion}

H-CogAff allows different representations of affective states. From the definitions in \cite{Sloman05:arch}, affective states are positive or negative: a negative state tends to cause avoidance of the state (e.g. pain), while a positive state tends to cause persistence of the state. Built on these definitions, emotions are defined by \cite{Sloman05:arch} as affective states that \q{interrupt or modulate the current process} (see also Simon \cite{simon67:control}). Modelling of emotions as interruptions is discussed elsewhere \cite{kennedy18:metacog} and is not a focus here, but it may become important in more advanced stages of behaviour change modelling.  We are interpreting an \q{affective state} as a {\it process} that evaluates a situation as positive or negative, which in turn causes a tendency towards action. This will be elaborated in Section \ref{sim}.

\section{SHARED CONCEPTS}\label{shared}

We have summarised two BCTs from health psychology (PRIME and CEOS) and introduced an approach to cognitive architectures in AI (H-CogAff). Table \ref{table2} identifies key concepts that are shared across these different models.
\begin{table}
\begin{center}
{\caption{Correspondences between concepts across theories.}\label{table2}}
\begin{tabular}{|l|l|l|}
\hline
H-CogAff & PRIME & CEOS  \\
\hline
\hline
Reactive & Motives, & Operational \\
layer & impulses, & system \\
 & responses & \\
\hline
Deliberative + & Plans, & Executive \\
metacognitive & evaluations, & system \\
layers & motive generation & \\
\hline
\end{tabular}
\end{center}
\end{table}
These are approximate correspondences. For instance, the CEOS operational level is highly associative and may respond to imagination (although it does not deliberately generate imaginary situations). It should be possible to integrate either of these BCTs into an architecture such as H-CogAff.

\subsection{Affective force}

Both CEOS and PRIME have the concept of \q{affective force}. This is mentioned explicitly in the CEOS theory:
\begin{quote}
\q{for ES goals to be pursued requires that the ES stimulates greater affective force and associated action tendencies within the OS supporting the goals than concurrent OS generated action tendencies} \cite{Borland17:CEOS}.   
\end{quote}
Similarly, PRIME principle 2 states that beliefs/intentions about good or bad must: \q{produce {\it sufficiently strong needs} at the moment of action} \cite{West07:PRIME}. 

The question is: how can executive processes generate affective force which is greater than the force that is activated in \q{the moment}? The following sections present our approach to the integration of affective force into cognitively rich simulated agents.

\section{AGENT SIMULATIONS}\label{sim}

To apply behaviour change theories to agent simulations, it is first necessary to define the components of a simulation. Cognitively rich agent simulations can be specified using the following components:
\begin{enumerate}
\item 
Scenario: a story involving agents and an environment.
\item
Ontology: what objects and agents exist? what are their relations? These can be derived from the scenario.
\item
Starting state: what does the world look like initially?
\item
Models (or theories) - in particular:
\begin{enumerate}
\item 
How does the world change? (e.g. physical models)
\item
How do agents make decisions and act? This is where the cognitive architecture and behaviour change theories are applied (e.g. H-CogAff + CEOS)
\end{enumerate}
\end{enumerate}

\subsection{Example Scenarios}

To specify problems that need to be solved by an agent, we define two example behaviour change scenarios in the form of simple stories as follows:
\begin{itemize}
\item
Scenario 1: Tom has decided to stop smoking and has so far succeeded. However, after a row with a colleague at the office, he feels angry and disappointed. These negative emotions interfere with his resolve not to smoke and he feels that a cigarette can help him calm down. To counteract these emotions, he thinks about the long-term advantages of persisting as a non-smoker and the need to put into perspective the relatively minor issue of the office confrontation.
\item
Scenario 2: Jane plans to cut down on sugar. However, an office colleague has brought some cake and has invited everyone to a birthday celebration. Jane finds it difficult to resist. After some consideration, she decides to join her colleagues, but she will explain that she wants to reduce her sugar intake and will bring some fruit instead. 
\end{itemize}
Both these scenarios are examples of a pattern where a goal is to be achieved, but in an adverse situation, opposing motives are generated because of needs that arise in that context. The person then takes countermeasures to strengthen their original resolve and to counteract the unwanted motive. These components of each scenario are shown in Table \ref{table1}.
\begin{table}
\begin{center}
{\caption{Details of each scenario.}\label{table1}}
\begin{tabular}{|l|l|l|l|l|}
\hline
Scenario & Goal & Adverse & Activated & Counter- \\
 & & situation & motive & measure
\\
\hline
\hline
1 & stop & row with & need for & situation \\
 & smoking & colleague & relaxation & redescription
\\
\hline
2 & reduce & office & want to be & imaginative\\
 & sugar & cake & friendly & replanning 
\\
\hline
\end{tabular}
\end{center}
\end{table}

\subsection{Affective agents}

In order to integrate \q{affective force} into a cognitive agent, it is first necessary to represent affect in an agent architecture.  As a working definition, we are defining an {\it affective process} as a sequence involving the following steps, based on Gross's process model \cite{Gross07:emotion}.
\begin{enumerate}
\item 
direct attention to a situation
\item
evaluate the situation - positively or negatively (possibly leading to further attention focus)
\item
prepare for action (which may lead to further evaluation of potential actions)
\end{enumerate}
There are two important concepts: {\it evaluation} and {\it action preparation}. The simplest kind of evaluation generates a positive or negative result (a neutral evaluation would not be affective). Evaluation fits into the more general concept of {\it appraisal} \cite{Scherer99:appraisal,Ortony88:OCC}, which can involve complex reasoning about why a situation may be an opportunity or threat. It is possible to have competing appraisals, where the same situation is described as negative using one interpretation and positive using another.

The importance of action preparation has been emphasised by Frijda \cite{frijda87:cog}. We are defining such as process as follows:
\begin{enumerate}
\item
Determine what states are desirable (what would be an improvement?)
\item
What desires should be pursued as goals? This involves reasoning about what is achievable, or compatible with other goals.
\item
What goals or actions should be chosen right now?
\end{enumerate}
These steps are primarily deliberative but can involve reactive layers. For example, step 1 might be the top-down (effortful) consideration of a hypothetical option or it might involve a fast associative process (e.g. imagining food because of hunger). Similarly steps 2 and 3 need not be deliberative, but include automatic learned sequences (e.g. avoiding a speeding vehicle). 

The action preparation stage will normally be interleaved with the evaluation and attention steps in the process model so that the process is iterative. The concept of \q{motive} can be simulated by this process. 

\subsection{Competing affective processes}
Scenarios 1 and 2 can be understood using competing affective processes involving iterative evaluation and action preparation. Table \ref{table2} shows example sequences for both scenarios. In both cases, the counteracting process starts to take effect after the second evaluation step.
\begin{table}
\begin{center}
{\caption{Affective processes for the non-smoking scenario}\label{table2}}
\begin{tabular}{|l|l|l|l|l|}
\hline
Scenario 1   & Eval1 &  Act1 & Eval2 & Act2 \\
\hline
\hline
 Proc1 &  bad mood & cigarette & calming & plan to smoke \\
\hline
 Proc2 & n/a & n/a & failure & avoid smoking \\
\hline
\end{tabular}
\end{center}
\end{table}
For each scenario, the two affective processes are labeled \q{proc1} and \q{proc2} respectively. In the non-smoking scenario, proc1 is the disruptive process that evaluates the current situation as negative (eval1: \q{bad mood}). As the first step in its action preparation, it proposes an improved state (act1: \q{have a cigarette}). The proposed action (smoking) is evaluated as positive because it is believed to improve mood (eval2). Then the next step in the action preparation is to form a goal to smoke (act2). The competing affective process (proc2) is the one that evaluates the situation of \q{having a cigarette} as negative because it conflicts with the goal of being a non-smoker. Proc2 needs to have greater affective force than proc1.
\begin{table}
\begin{center}
{\caption{Affective processes for the office cake scenario}\label{table2}}
\begin{tabular}{|l|l|l|l|l|}
\hline
Scenario 2   & Eval1 &  Act1 & Eval2 & Act2 \\
\hline
\hline
Proc1 &  problematic & make an & being & have cake \\
   &   & exception & friendly &  \\
\hline
Proc2 & n/a & n/a & giving in & plan alternative \\
\hline
\end{tabular}
\end{center}
\end{table}
For the office cake scenario, proc1 begins with an evaluation of a situation: eval1 (\q{problematic}) because it interferes with the initial goal (reduce sugar). An action preparation (act1) proposes a potential solution (\q{make an exception}). This plan is evaluated positively as \q{being friendly} (eval2), which in turn leads to an intention to execute it (act2). Proc2 then counteracts this by proposing an alternative plan.  

\subsection{Role of metacognition in affective processes}

Metacognition is not only \q{thinking about thinking}; it can also involve thinking about motivation and emotion (e.g. \q{I'm being affected by anger}). In the two scenarios in Table \ref{table2}, the counteracting process (proc2) may be metacognitive. For example, in the non-smoking scenario, a metacognitive version of proc2 might detect that the reasoning process underpinning \q{smoking is calming} is biased because it is affected by emotion. In particular, it is forgetting the initial goal and overlooking the importance given to this goal originally. In other words, a metacognitive process evaluates a reasoning process, not merely a potential action or state. This has advantages because the agent may be able to identify and explain the reason for the current tendency to smoke, which allows the possibility of finding a more satisfactory reasoning process, and in turn lead to more positive emotions.

To understand how metacognition can work in detail, it is necessary to consider how the simulation will be implemented. 

\section{IMPLEMENTING A PROTOTYPE}\label{prot}

Current work is investigating strategies for implementing an example of a simulated agent which is acting according to a BCT such as PRIME and demonstrates self-regulation using metacognition, particularly involving the concept of affective force.

\subsection{Metacognitive monitoring}

To implement metacognitive monitoring, the process being monitored needs to leave a trace of its reasoning \cite{cox07:explain}. The trace acts like a memory or history of mental events (such as decisions made at different points). This is then analysed by the metacognition.  For example, the following key variables can be included in the trace:
\begin{itemize}
\item
changes in attention focus.
\item
changes in beliefs and evaluations (appraisals).
\item
changes in deliberation state: (a) current goal and options; (b) which option was selected and why.
\end{itemize}
Changes in these variables will occur naturally as the agent carries out a task.  A change indicates a problem if any key variable values are inconsistent with those implied by the initial goal. These might include new desirable states that are inconsistent. For example, if the potential action of \q{smoking} is suddenly evaluated positively (a cigarette would be good) instead of negatively, the evaluation would be recognised as a problem (because the  initial goal implies that smoking is bad). This is a process of logical consistency checking, similar to belief revision but with a different purpose. For belief revision, if the observations are inconsistent with beliefs, the beliefs need to change; for goal achievement, the subsequent desirable states or subgoals need to be consistent with the initial goal; otherwise they are discarded. (This is the contrast between \q{desire-like} and \q{belief-like} states mentioned in \cite{Sloman05:arch}). In addition to evaluations, other aspects of an affective process can be monitored: for example, is the action preparation or attention focus consistent with the goal?

\subsection{Scenarios}

Metacognition can be more easily implemented if there is a well-defined task where progress can be monitored. The two scenarios (1 and 2) are not easy to translate into this kind of task. It is possible, however, to identify the important patterns in the scenarios and apply them to a simplified world. Instead of a goal such as \q{not smoking}, the goal needs to be a task whose progress can be monitored. We are currently investigating a spatial scenario where an agent has to tidy a room in which various objects are scattered about. This can be converted into a behaviour change scenario by introducing difficulties into the virtual world that can activate unwanted motivations. For example, in the room tidying task, the goal can be a state where all books are in the correct position in the shelves, and toys are put away in a box. \\ \\
\noindent
This modified scenario consists of three processes instead of two:
\begin{itemize}
\item
proc0: initial process which performs task
\item
proc1: conflicting motive caused by a difficulty
\item
proc2: countermeasure to restore work on task
\end{itemize}
For example, in the room tidying task, the original goal-seeking activity (proc0) can be defined as an affective process, since untidiness can be evaluated as unpleasant or ugly, while a future tidy room is satisfying. The general goal of the agent is to be tidy and disciplined (like \q{non-smoker}, only positively defined). A difficulty can arise if, for example, the shelves in the room are broken and the books cannot be placed on them. A negative reaction (proc1) is to want to give up and leave the room in an untidy state (because trying to fix shelves is unpleasant and unexpected). The countermeasure involves finding a way of stacking the books temporarily on a table, so that the room still looks tidy.

\section{CHALLENGES AND FUTURE WORK}\label{challenges}

There are significant research challenges, which are identified below.

\subsection{Affective force and reasoning}

An important aspect of \q{affective force} is providing {\it reasons} to do things. For example, in the non-smoking scenario, the counteracting process uses the reason of \q{having made a commitment}. For this purpose, it is possible to represent the competing affective processes as {\it arguments} for and against an action. This is similar to what happens in real life. In the non-smoking scenario, the argument in favour of smoking would be: \q{a cigarette would be calming right now}, while the counter argument would say: \q{no, smoking now would break an earlier commitment}.

This kind of approach is already used in the agent decision-making model of \cite{Fox13:canonical}), which constructs reasons for and against an option, and where reasons can have numerical weights. The final decision recommendation is made using an algorithm for aggregating the different reasons for and against each option. A simple numerical solution would be to select the option with most arguments in favour (and with highest total weight). However, interdependencies between arguments are possible, which may require logical reasoning.

\subsection{Metacognitive control}

Metacognitive control needs to generate alternative forms of reasoning. In the non-smoking scenario, the situation is re-described to highlight the importance of the initial goal in relation to the current difficulties. In the cake scenario, an alternative plan is generated. For early prototypes, these alternative forms of reasoning can be pre-programmed in advance as a kind of \q{library}, from which the metacognitive control can choose. In more advanced stages, learning and adaptation needs to be added. One approach is to use the critic-selector architecture of \cite{minsky07:book} as a framework for selecting different \q{ways to think}.

\subsection{Scenarios}

There needs to be a methodology for generating real-world scenarios. People who are attempting to achieve their own goals can be a valuable source of insights, particularly if they have failed a number of times and if they have experience using technology (such as a fitness or time-management app). Different classes of scenarios can be defined. For instance, we have only considered situations where a self-determined goal already exists; it also important to consider how such an intrinsic goal is formed.  

A methodology is also required for translating real-world scenarios into simplified visual tasks, so that the critical features of the original story are not lost. It is important to be aware of the original nuances and complexities even if they cannot all be reproduced in the simplified scenario.

\subsection{Pathway to real-world assistant agents}

In addition to gaining insights using simulation, our goal is also to find a pathway from autonomous simulated agents to real-world assistant agents. This requires an open source research infrastructure, which supports simulation and real-world prototyping on different levels, and which enables the building of knowledge-based assistant agents, some of which may be decision support. Although knowledge-based approaches are useful for transparency, they may be combined with other approaches, such as neural networks. The required research platform needs to support exploration of differing theories and definitions, without forcing a commitment to any particular one.

\bibliography{cog1}

\begin{thebibliography}{10}

\bibitem{Bandura91:SRT}
A.~Bandura, `{Social cognitive theory of self-regulation}', {\em Organizational
  Behavior and Human Decision Processes}, {\bf 50}(2),  248--287, (1991).

\bibitem{bello17:agency}
P.~Bello and W.~Bridewell, `{There is no Agency without Attention}', {\em AI
  Magazine}, {\bf 38}(4), (2017).

\bibitem{Borland17:CEOS}
R.~Borland, `{CEOS Theory: A Comprehensive Approach to Understanding Hard to
  Maintain Behaviour Change}', {\em Applied Psychology: Health and Wellbeing},
  {\bf 9},  3--35, (2017).

\bibitem{Carver82:control}
C.~S. Carver and M.~F. Scheier, `{Control Theory: A Useful Conceptual Framework
  for Personality-Social, Clinical, and Health Psychology}', {\em Psychological
  Bulletin}, {\bf 92}(1),  111--135, (1982).

\bibitem{cox07:explain}
M.~T. Cox, `Metareasoning, monitoring, and self-explanation', in {\em
  Proceedings of the First International Workshop on Metareasoning in
  Agent-based Systems at AAMAS-07}, (2007).

\bibitem{cox11:meta-intro}
M.~T. Cox and A.~Raja, `{Metareasoning: an Introduction}', in {\em
  Metareasoning: Thinking about Thinking}, eds., M.~T. Cox and A.~Raja,  3--14,
  MIT Press, (2011).

\bibitem{cox99:multistrategy}
M.~T. Cox and A.~Ram, `{Introspective multistrategy learning: On the
  construction of learning strategies}', {\em Artificial Intelligence}, {\bf
  112},  1--55, (1999).

\bibitem{Fleming17:serious}
Theresa~M. Fleming, Lynda Bavin, Karolina Stasiak, Eve Hermansson-Webb,
  Sally~N. Merry, Colleen Cheek, Mathijs Lucassen, Ho~Ming Lau, Britta
  Pollmuller, and Sarah Hetrick, `Serious games and gamification for mental
  health: Current status and promising directions', {\em Frontiers in
  Psychiatry}, {\bf 7}, (2017).

\bibitem{Fox13:canonical}
J.~Fox, R.~P. Cooper, and D.~W. Glasspool, `{A Canonical Theory of Dynamic
  Decision-Making}', {\em Frontiers in Psychology}, {\bf 4}(150), (2013).

\bibitem{frijda87:cog}
N.~H. Frijda, `Emotion, cognitive structure, and action tendency', {\em
  Cognition and Emotion}, {\bf 1}(1),  115--143, (1987).

\bibitem{Gollwitzer99:imp}
P.~M. Gollwitzer, `{Implementation intentions: Strong effects of simple
  plans}', {\em American Psychologist}, {\bf 54},  493--503, (1999).

\bibitem{Gross07:emotion}
J.~Gross and R.~Thompson, `{Emotion Regulation: Conceptual Foundations}', in
  {\em Handbook of Emotion Regulation}, ed., W~Gray, Guilford Publications, New
  York, (2007).

\bibitem{hudlicka07:reasons}
E~Hudlicka, `Reasons for emotions: modeling emotions in integrated cognitive
  systems', in {\em Integrated Models of Cognitive Systems}, ed., W.~Gray,
  1--37, Oxford University Press, New York, (2007).

\bibitem{hudlicka17:healthcare}
E~Hudlicka, `{Computational Modeling of Cognition - Emotion Interactions:
  Theoretical and Practical Relevance for Behavioral Healthcare}', in {\em
  Emotions and Affect in Human Factors and Human-Computer Interaction}, ed.,
  M.~P. Jeon,  383--436, Academic Press, Orlando, FL, (2017).

\bibitem{Broekens16:appraisal}
Broekens J., Hudlicka E., and Bidarra R., `{Emotional Appraisal Engines for
  Games}', in {\em {Socio-Affective Computing, vol 4.}}, eds., Karpouzis K. and
  Yannakakis G., Springer, Cham, (2016).

\bibitem{kennedy18:metacog}
C.~M. Kennedy, `{Computational Modelling of Metacognition in Emotion
  Regulation}', in {\em 8th Workshop on Emotion and Computing at the German
  Conference on AI (KI-2018)}, Berlin, Germany, (2018).

\bibitem{Liu14:KBS}
S.~Liu and P.~Zarate, `{Knowledge Based Decision Support Systems: A Survey on
  Technologies and Applications Domains}', in {\em Joint International
  Conference on Group Decision and Negotiation (GDN-2014), Lecture Notes in
  Business Information Processing}, volume 180, pp. 62--72, Toulouse, France,
  (2014).

\bibitem{Marsella07:EMA}
S.~C. Marsella and J.~Gratch, `{EMA: A process model of appraisal dynamics}',
  {\em Cognitive Systems Research}, {\bf 10},  70--90, (2007).

\bibitem{Michie14:ABC}
S.~Michie, R.~West, Campbell R., J.~Brown, and H.~Gainforth, {\em {An ABC of
  Behaviour Change Theories}}, Silverback Publishing, 2014.

\bibitem{minsky07:book}
M~Minsky, {\em The Emotion Machine: Commonsense thinking, artificial
  intelligence, and the future of the human mind}, Simon and Shuster, New York,
  2007.

\bibitem{Ortony88:OCC}
A.~Ortony, G.~Clore, and A.~Collins, {\em The Cognitive Structure of Emotions},
  Cambridge University Press, Cambridge, MA, 1988.

\bibitem{Popescu14:GAM}
A.~Popescu, J.~Broekens, and M.~van Someren, `{GAMYGDALA: An Emotion Engine for
  Games}', {\em IEEE Transactions on Affective Computing}, {\bf 5}(1),  32--44,
  (2014).

\bibitem{Ryan02:SDT}
R.~M. Ryan and E.~L. Deci, `An overview of self-determination theory', in {\em
  Handbook of self-determination research}, eds., Deci~E. L. and R.~M. Ryan,
  University of Rochester Press, Rochester, NY, (2002).

\bibitem{Scherer99:appraisal}
K.~R. Scherer, `Appraisal theory', in {\em Handbook of Cognition and Emotion},
  eds., T.~Dalgleish and M.~Power, John Wiley and Sons, Chichester, UK, (1999).

\bibitem{Schuller13:pervasive}
B.~Schuller, I.~Dunwell, F.~Weninger, and L.~Paletta, `{Pervasive Serious
  Gaming for Behavior Change - The State of Play}', {\em IEEE Pervasive
  Computing}, {\bf 12}(3), (2013).

\bibitem{simon67:control}
H.~A. Simon, `Motivational and emotional controls of cognition', {\em
  Psychological Review}, {\bf 74}(1),  29--39, (1967).

\bibitem{Sloman05:arch}
A.~Sloman, R.~Chrisley, and M.~Scheutz, `The architectural basis of affective
  states and processes', in {\em Who Needs Emotions?}, eds., J.-M. Fellous and
  M.A Arbib, Oxford University Press, New York, (2005).

\bibitem{Strack04:RIM}
F.~Strack and R.~Deutch, `{Reflective and Impulsive Determinants of Social
  Behaviour}', {\em Personality and Social Psychology Review}, {\bf 8}(3),
  (2004).

\bibitem{West07:PRIME}
R.~West, `{The PRIME Theory of motivation as a possible foundation for
  addiction treatment}', in {\em Drug Addiction Treatment in the 21st Century:
  Science and Policy Issues}, eds., J.~Henningfield, P.~Santora, and R.~West,
  John's Hopkins University Press, Baltimore, (2007).

\end{thebibliography}

\end{document}